\documentclass[amsmath,aps,prb,twocolumn,longbibliography,superscriptaddress]{revtex4-2}
\usepackage{graphicx}
\usepackage{amsmath}
\usepackage{amssymb}
\usepackage{bm}
\usepackage{enumerate}
\usepackage{xfrac}
\usepackage[caption=false]{subfig}
\usepackage[normalem]{ulem}
\usepackage[usenames,dvipsnames]{xcolor}
\usepackage{txfonts}
\usepackage[mathscr]{euscript}
\usepackage[pdftex,
pdflang={en-US},
colorlinks=true,
pdfauthor={Pontus Laurell},
pdftitle={Spin dynamics of the generalized quantum spin compass chain}]{hyperref}
\usepackage{verbatim}
\usepackage{ulem}

\usepackage{subdepth}

\begin{document}
	



\title{Spin dynamics of the generalized quantum spin compass chain}
\author{Pontus Laurell}
\email{plaurell@utk.edu}
\affiliation{Department of Physics and Astronomy, University of Tennessee, Knoxville, Tennessee 37996, USA}
\affiliation{Computational Sciences and Engineering Division, Oak Ridge National Laboratory, Oak Ridge, Tennessee 37831, USA}
\author{Gonzalo Alvarez}
\affiliation{Computational Sciences and Engineering Division, Oak Ridge National Laboratory, Oak Ridge, Tennessee 37831, USA}
\author{Elbio Dagotto}
\affiliation{Department of Physics and Astronomy, University of Tennessee, Knoxville, Tennessee 37996, USA}
\affiliation{Materials Science and Technology Division, Oak Ridge National Laboratory, Oak Ridge, Tennessee 37831, USA}

\date{\today}

\begin{abstract}
	We calculate the dynamical spin structure factor of the generalized spin-$1/2$ compass spin chain using the density matrix renormalization group. The model, also known as the twisted Kitaev spin chain, was recently proposed to be relevant for the description of the spin chain compound CoNb$_2$O$_6$. It  features bond-dependent interactions and interpolates between an Ising chain and a one-dimensional variant of Kitaev's honeycomb spin model. The structure factor, in turn, is found to interpolate from gapped and non-dispersive in the Ising limit to gapless with non-trivial continua in the Kitaev limit. In particular, the component of the structure factor perpendicular to the Ising directions changes abruptly at the Kitaev point into a dispersionless continuum due to the emergence of an extensive groundstate degeneracy. We show this continuum is consistent with analytical Jordan-Wigner results. We also discuss implications for future inelastic scattering experiments and applications to materials, particularly CoNb$_2$O$_6$.
\end{abstract}

\maketitle

\section{Introduction}
Orbital physics in Mott insulators can lead to a wide range of important phenomena \cite{Khaliullin2005, Georges2013, RevModPhys.87.1, Rau2016, doi:10.1021/acs.chemrev.0c00579} including dimensionality reduction \cite{doi:10.1021/acs.chemrev.0c00579}, orbital-selective Mott phases \cite{PhysRevLett.92.216402, PhysRevB.72.205124, PhysRevLett.123.027203}, and, in the presence of spin-orbit coupling, bond-dependent magnetic exchange interactions \cite{Khaliullin2005, PhysRevLett.102.017205, RevModPhys.87.1}. The latter feature dramatically in \emph{compass} models \cite{RevModPhys.87.1}, with Ising interactions along specific spin-space directions depending on the spatial bond  direction. A famous example is Kitaev's honeycomb spin model \cite{Kitaev2006}, which realizes a quantum spin liquid ground state. Its possible material realizations have been the subject of intense research recently \cite{Takagi2019, doi:10.7566/JPSJ.89.012002, Trebst2022}.

Another intriguing example is the 1D quantum compass model (QCM) with alternating $S_i^xS_{i+1}^x$ and $S_{i+1}^yS_{i+2}^y$ interactions for different bonds along the chain direction \cite{PhysRevB.75.134415, PhysRevB.78.184406}, which provides an exactly solvable model presenting a quantum multicritical point \cite{PhysRevB.79.224424, PhysRevB.80.174417} in extended models. The QCM can be viewed as arising from orbital order in systems of weakly interacting zigzag chains \cite{PhysRevB.89.104425}, or simply as a 1D version of Kitaev's honeycomb model: a Kitaev spin chain. Chain and ladder versions of the Kitaev honeycomb model and its extensions (including e.g. Heisenberg and off-diagonal Gamma interactions) have been studied theoretically \cite{PhysRevB.79.224408, PhysRevB.86.205412, PhysRevB.96.041115, PhysRevB.96.205109, Agrapidis2018, EfthimiaAgrapidis2018, Agrapidis2018a, PhysRevA.98.052303, PhysRevB.99.195112, PhysRevB.99.195112, PhysRevB.99.205129, PhysRevB.99.224418, PhysRevLett.124.147205, PhysRevA.102.012406,  PhysRevResearch.2.033268, PhysRevB.102.144437,  PhysRevResearch.3.033048, PhysRevX.11.011013, PhysRevB.103.054437, PhysRevB.103.195102, PhysRevResearch.4.013205, PhysRevResearch.4.013103, PhysRevB.105.094432, PhysRevB.106.064425, PhysRevResearch.5.L012027}, mostly for their tractability and potential realizations in engineered chains \cite{https://doi.org/10.1002/adma.201603798}. It was also proposed that charge order in K-intercalated RuCl$_3$ may lead to effective Kitaev-Heisenberg chains \cite{Agrapidis2018, Agrapidis2018a}, but a different charge order was found in experiments \cite{PhysRevMaterials.1.052001}.

Given the above information, zigzag chains appear to be the most promising way towards such 1D Kitaev-like models in materials. Due to the variability of bond angles and lattice distortions, it is natural to consider a generalized compass model (GCM) \cite{PhysRevB.89.104425, Morris2021},
\begin{align}
	H	&=	-K\sum_{i=0}^{L/2-1} \left( \tau_{2i}^{\hat{n}_1} \tau_{2i+1}^{\hat{n}_1} + \tau_{2i+1}^{\hat{n}_2} \tau_{2i+2}^{\hat{n}_2}\right),	\label{eq:compass}
\end{align}
where $\tau_i^{\hat{n}_j}	=\hat{n}_j \cdot \vec{\tau_i}$ 
is the projection of the pseudospin Pauli operator vector on site $i$ onto the bond-dependent Ising direction $\hat{n}_j$. Using a coordinate system where the two axes $\hat{n}_1$ and $\hat{n}_2$ lie in a plane, we allow the angle $2\theta$ between $\hat{n}_1$ and $\hat{n}_2$ to vary continuously. At $\theta=0, \pi/2$ the Ising chain is recovered, while $\theta=\pi/4$ yields the QCM \footnote{This is the same convention for the angle as in Ref.~\cite{Morris2021} and half the angle of Ref.~\cite{PhysRevB.89.104425}.}, which was solved in the seventies as a special case of the alternating $XY$ model \cite{PERK1975319}. 
The interpolation between Ising and Kitaev spin chains motivated Morris \emph{et al.}~\cite{Morris2021} to introduce ``twisted Kitaev spin chain'' as an alternate name for the GCM away from these limits. 
They also proposed the Hamiltonian~\eqref{eq:compass} as a description of long-distance properties in the zigzag chain material CoNb$_2$O$_6$ \cite{Morris2021}, which is commonly considered the best known realization of the ferromagnetic (FM) transverse-field spin-1/2 Ising chain due to its observed field-induced criticality \cite{Coldea2010, Lee2010, PhysRevB.83.020407, PhysRevB.102.104431, PhysRevX.12.021020}. 
The description as a pure FM Ising chain is, however, insufficient to explain the zero-field behavior, the description of which motivates considering bond-dependent interactions  \cite{doi:10.1073/pnas.2007986117, Morris2021}.

\begin{figure}
	\includegraphics[width=\columnwidth]{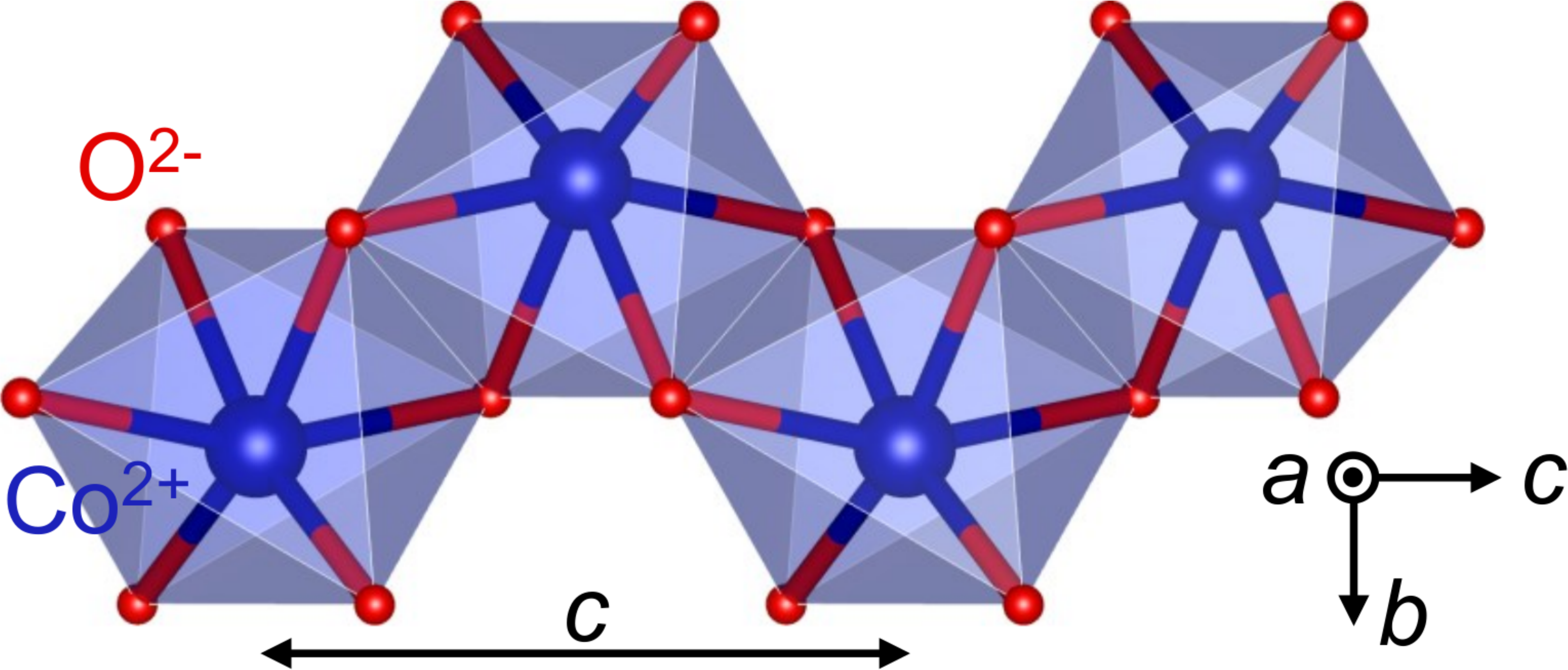}
	\caption{\label{fig:zigzagchain}Zigzag chain in CoNb$_2$O$_6$ featuring a two-site unit cell with lattice constant $c$ along the chain direction. The lattice symmetry allows for different interactions between spins along the two distinct bonds. Similar bond-dependent interactions may emerge also in other zigzag chain systems with specific electron configurations.
	}
\end{figure}
What would originate such interactions in CoNb$_2$O$_6$? Their Co$^{2+}$ ions are surrounded by oxygen octahedral cages and form zigzag chains along the $c$ axis; see Fig.~\ref{fig:zigzagchain}. Hund's coupling favors a high-spin $d^7$ configuration ($t_{2g}^5e_g^2$), which may be viewed as a $S=3/2$, $L=1$ state. Spin-orbit coupling then splits the energy levels further, resulting in a pseudospin-1/2 ground state Kramers doublet, just as in proposals for Kitaev physics in honeycomb cobaltate systems \cite{PhysRevB.97.014407, Kim2022}. Although CoNb$_2$O$_6$ is not a honeycomb system, its symmetry permits identification of two alternating Ising directions \cite{Morris2021}. 
Distortion of the octahedra splits the energy levels further, but the ground state Kramers doublet remains \cite{PhysRevB.105.224421}. 
We note that the GCM, Eq.~\eqref{eq:compass}, is general and not restricted to materials such as CoNb$_2$O$_6$. 
It may also emerge in $d^9$, high-spin $d^4$, and low-spin $d^7$ configurations, where the $e_g$ orbital degree of freedom replaces the Kramers doublet degree of freedom \cite{PhysRevB.89.104425}. Further potential applications include Co zigzag chains on surfaces \cite{Dup__2015} and quantum simulation in optical lattices \cite{Simon2011, PhysRevB.86.155159}. 

Since Eq.~\eqref{eq:compass} and variations of the model are exactly solvable using Jordan-Wigner fermions \cite{Jordan1928, Lieb1961}, many properties have been studied. These include ground state properties \cite{PhysRevB.75.134415, PhysRevB.89.104425, PhysRevB.78.184406, Brzezicki2009, PhysRevB.84.035112, Motamedifar2011,  PhysRevB.90.094413, Qiu_2016, PhysRevB.100.085130} , thermodynamic properties \cite{Jafari2012, PhysRevB.85.184422, PhysRevB.89.104425}, and aspects of quantum quench dynamics \cite{Jafari_2016, PhysRevLett.118.015701, Jafari2019}. Numerical results were also reported in Refs.~\cite{PhysRevB.80.174417, Motamedifar2011, Mahdavifar2010} using Lanczos exact diagonalization and Ref.~\cite{PhysRevB.85.184422} using matrix product state methods. However, to the best of our knowledge, the full dynamical spin structure factor $S(k,\omega)$ has not yet been studied except in the Ising limit, although time-dependent results for the spin dynamics of the QCM 
were obtained analytically for spin components in the plane spanned by $\hat{n}_{1}$ and $\hat{n}_2\perp \hat{n}_1$ \cite{PERK1977265} and for spin components transverse to the same plane \cite{Perk1977}\footnote{The QCM possesses a factorization property \cite{PERK1977265}, through which its spin dynamics can be related to two transverse-field Ising chain correlation functions, for which bulk results are available \cite{Perk2009}.}. The goal of the present paper is thus to study the \emph{frequency-dependent dynamics} at zero temperature and as function of the angle $\theta$.

Using the density matrix renormalization group (DMRG) \cite{PhysRevLett.69.2863, PhysRevB.48.10345} we obtain all components of $S(k,\omega)$ as a function of the angle $\theta$. The spectra interpolate from gapped and non-dispersive at the Ising points towards a gapless continuum as the Kitaev point is approached, with gapped and dispersive behavior in-between. There are abrupt qualitative changes in the spectra at the Kitaev point, related to an underlying macroscopic degeneracy. In particular, the transverse component  $S^{yy}(k,\omega)$ becomes gapless and dispersionless in the Kitaev limit. 
These spectral features are understood via the Jordan-Wigner ground state solution. 
Our $S(k,\omega)$ results can help the design and interpretation of future experiments employing, for example, inelastic neutron scattering (INS) or resonant inelastic xray scattering (RIXS) techniques.

The paper is organized as follows. Sec.~\ref{sec:model} introduces global coordinate systems for Eq.~\eqref{eq:compass} to interrelate the conventions of Refs.~\cite{PhysRevB.89.104425, Morris2021}. We review relevant Jordan-Wigner results in Sec.~\ref{sec:JW} and describe the numerical methods in Sec.~\ref{sec:methods}. We present our results in Sec.~\ref{sec:results}, discuss their consequences and summarize the conclusions in Sec.~\ref{sec:discussion}. 
A derivation of the dispersionless continuum at the Kitaev point is provided in Appendix~\ref{app:dispersionlesscontinuum}.

\section{Coordinate systems}\label{sec:model}
For concreteness, we first consider the application of Eq.~\eqref{eq:compass} to CoNb$_2$O$_6$. The crystal structure features zigzag chains along the crystallographic $c$ axis as shown in Fig.~\ref{fig:zigzagchain}, in which the two Ising directions are constrained by symmetry to be related by a twofold rotation symmetry about $b$, $C_2^b$. Following Morris \emph{et al.} \cite{Morris2021} we use a global $xyz$ coordinate system where two Ising directions $\hat{n}_1,\hat{n}_2$ define the $xz$-plane. This is done by choosing $\hat{x}$ parallel with the $b$ axis, and $\hat{z}$ such that it bisects the angle $2\theta \approx 34^\circ$ between $\hat{n}_1$ and $\hat{n}_2$ and is at an angle $\phi\approx 31^\circ$ to the $c$ axis. The first Ising axis can be taken as $\hat{n}_1=\left(\sin \theta,0,\cos\theta\right)$, with $\hat{n}_2$ fixed by $C_2^b$ symmetry.

Substituting the $\hat{n}_j$ 
into Eq.~\eqref{eq:compass}, transforming to pseudospin-1/2 operators $S_i^a=\tau_i^a/2$ and defining $\tilde{K}=4K$ one obtains 
\begin{align}
H_1	&=	-\tilde{K} \sum_i \left[ \cos^2\left(\theta \right) S_i^z S_{i+1}^z + \sin^2\left(\theta\right) S_i^x S_{i+1}^x \right.\nonumber\\
&\left. + \frac{\sin\left(2\theta\right)}{2} \left( -1\right)^i \left( S_i^x S_{i+1}^z + S_i^z S_{i+1}^x\right) \right],	\label{eq:pauliham:morris}
\end{align}
as in Ref.~\cite{Morris2021}. In the absence of magnetic fields there is a twofold ground state degeneracy due to invariance under spin rotations around $\hat{y}$ by $\pi$. We call this the Ising-like coordinate system because the Ising nature of the Hamiltonian is manifest at $\theta=0,\pi/2$. However, since the bond alternation is in the symmetric off-diagonal (or $\Gamma$) terms, the Kitaev nature at $\pi/4$ is obscured:
\begin{align}
	H_1^{\theta=\pi/4}	&=	-\frac{\tilde{K}}{2} \sum_i \left[ S_i^zS_{i+1}^z + S_i^x S_{i+1}^x + \left( -1\right)^i \left( S_i^xS_{i+1}^z + S_i^z S_{i+1}^x \right)\right].
\end{align}

The connection to Kitaev or compass physics becomes clearer by canonically transforming to an alternate coordinate system $(x^\prime y^\prime z^\prime)$ by a $\pi/4$ counterclockwise rotation around $-\hat{y}$. In this Kitaev-like coordinate system the bond-alternation is moved to the Ising terms,
\begin{align}
&H_2	=	- \frac{\tilde{K}}{2}\sum_i \left\{ \left[1- \left(-1\right)^i \sin\left(2\theta\right)\right] S_{i}^{x^\prime} S_{i+1}^{x^\prime} \right.	\label{eq:pauliham:kitaevlike}\\
	&\left. + \left[1+\left(-1\right)^i\sin\left(2\theta\right)\right] S_{i}^{z^\prime} S_{i+1}^{z^\prime}  - \cos\left( 2\theta\right) \left[ S_i^{x^\prime} S_{i+1}^{z^\prime} + S_i^{z'}\nonumber S_{i+1}^{x'} \right]\right\},	
\end{align}
making the Kitaev nature manifest at $\theta=\pi/4$. The drawback is that the Ising nature at $\theta=0,\pi/2$ is now obscured, where the Hamiltonian takes the form of an X'Y' model with a  $\Gamma$ interaction term. We will report our spin dynamics results in the Ising-like coordinate system, both because of its established connection to experimentally relevant systems and because the rotation to the Kitaev-like coordinate system generically induces off-diagonal $S^{x^\prime z^\prime/z^\prime x^\prime}(k,\omega)$ correlations, which can be significant. 

Finally, to connect with prior Jordan-Wigner analyses of the GCM it is convenient to apply a $\pi/2$ spin rotation about $\hat{x}$,
\begin{align}
&S^x	\rightarrow \tilde{S}^x,	\quad S^y	\rightarrow \tilde{S}^z,\quad S^z	\rightarrow -\tilde{S}^y,	\label{eq:spinrot}
\end{align}
to Eq.~\eqref{eq:pauliham:morris}, yielding
\begin{align}
H_3	&=	-\tilde{K}\sum_i \left[ \sin^2 \left( \theta\right) \tilde{S}_i^x \tilde{S}_{i+1}^x + \cos^2 \left( \theta\right) \tilde{S}_i^y \tilde{S}_{i+1}^y \right.\nonumber\\
&\left. - \frac{\sin\left( 2\theta\right)}{2} \left( -1\right)^i \left( \tilde{S}_i^x \tilde{S}_{i+1}^y + \tilde{S}_i^y \tilde{S}_{i+1}^x \right)\right].	\label{eq:spinham:rotated}
\end{align}
In the following we will use $H_3$ in the discussion of the Jordan-Wigner solution, but present spin dynamics results in the coordinate system of $H_1$. This approach gives both a concrete connection to CoNb$_2$O$_6$ and similar systems, and increased numerical efficiency from working with real-valued Hamiltonians.

\section{Jordan-Wigner solution}\label{sec:JW}
We review here aspects of the exact solution of the model in the Jordan-Wigner formalism \cite{Jordan1928, Lieb1961}, following mainly Refs.~\cite{PhysRevB.89.104425, Jafari_2016}. Introducing the standard transformation
\begin{align}
\tilde{S}_i^+	=	\tilde{S}_i^x + i\tilde{S}_i^y	&= c_i^\dagger \exp \left[ i\pi \sum_{j=1}^{i-1} c_j^\dagger c_j \right],\\
\tilde{S}_i^-	=	\tilde{S}_i^x - i\tilde{S}_i^y	&= \exp \left[ -i\pi \sum_{j=1}^{i-1} c_j^\dagger c_j \right] c_i,	\\
\tilde{S}_i^z	&=	c_i^\dagger c_i - \frac{1}{2},
\end{align}
where $\left\{ c_i, c_j^\dagger \right\} = \delta_{i,j}$, Eq.~\eqref{eq:spinham:rotated} is recast in terms of spinless fermions,
\begin{align}
H_3	&=	-K	\sum_{i=1}^L \left[ c_i^\dagger c_{i+1} + \mathrm{H.c.} \right] \nonumber\\
&+ K\sum_{i=1}^{L/2}	\left[ c_{2i}^\dagger c_{2i+1}^\dagger e^{-i2\theta} + c_{2i+1}^\dagger c_{2i+2}^\dagger e^{i2\theta} + \mathrm{H.c.} \right],	\label{eq:hamJW}
\end{align}
where $L$ is the length of the chain, $L/2$ is the number of unit cells, and $\mathrm{H.c.}$ denotes Hermitian conjugate. We adopt a periodic Fourier convention with
\begin{align}
c_{2j-1}	&=	{\sqrt{\frac{2}{L}}}	\sum_ke^{-ikj} a_k,		\quad c_{2j}		=	{\sqrt{\frac{2}{L}}} \sum_k e^{-ikj} b_k,	\label{eq:FTconv}
\end{align}
and momenta given by
\begin{align}
k	&=	\frac{2n\pi}{L},	\quad	n=-\left( \frac{L}{2}-1\right), - \left(\frac{L}{2}-3\right), \dots ,\left(\frac{L}{2}-1\right).
\end{align}

Following the Fourier transform, Eq.~\eqref{eq:hamJW} is rewritten in a symmetrized Bogoliubov-de Gennes form,
\begin{align}
	\mathcal{H}	&=	\frac{1}{2}\sum_k \Gamma_k^\dagger h(k) \Gamma_k,	\quad \Gamma_k^\dagger =\left( a_k^\dagger, a_{-k}, b_k^\dagger, b_{-k}\right),	\label{eq:symmetrizedbdg:ham}
\end{align}
where
\begin{align}
h(k)	&=	%
\begin{pmatrix}
0 						& 0 										& A_k						&	P_k + Q_k\\
0 						& 0 										& -\left( P_k - Q_k\right)	&	-A_k\\
A_k^\star				&	-\left( P_k^\star - Q_k^\star \right)	&	0						&	0\\
P_k^\star + Q_k^\star	&	-A_k^\star								&	0						&	0
\end{pmatrix}	\label{eq:ham:JW:BdG:matrix}
\end{align}
and
\begin{align}
A_k	&=	-K \left( 1+ e^{ik}\right),	\\
P_k	&=	K\cos \left( 2\theta\right) \left( 1-e^{ik}\right),	\quad	Q_k	=	iK\sin \left( 2\theta\right) \left( 1 + e^{ik} \right).
\end{align}

\begin{figure}
	\includegraphics[width=\columnwidth]{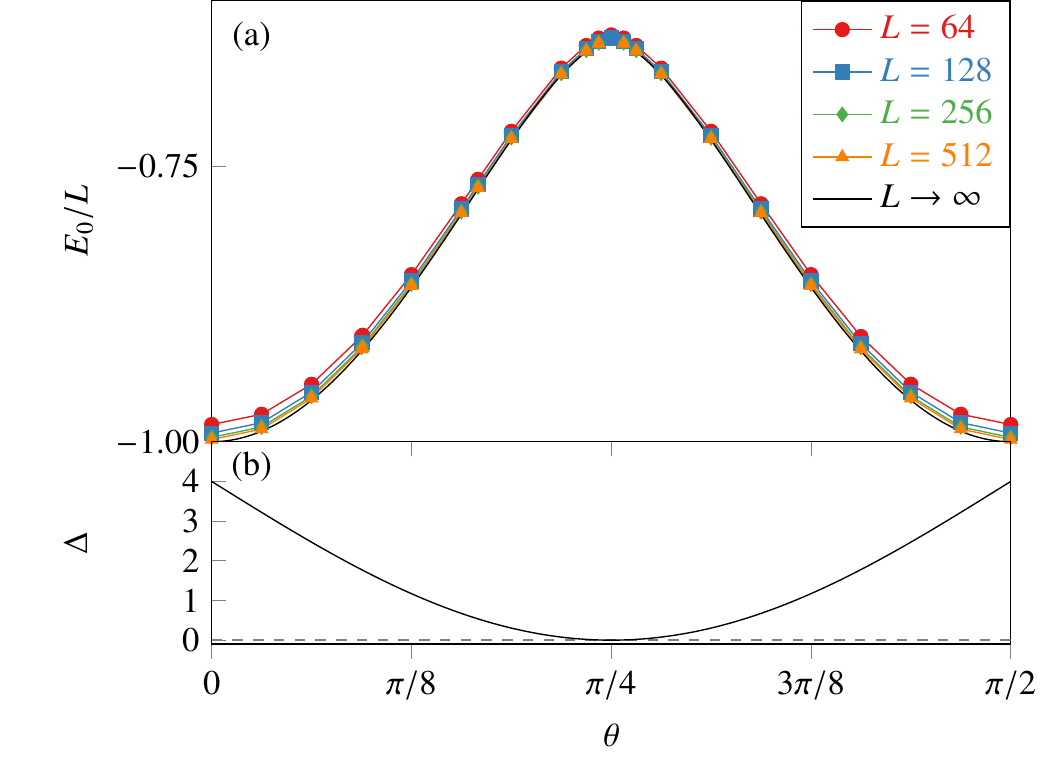}
	\caption{\label{fig:energy:FM}(a) Groundstate energy in units where $\left| K\right|=1$. The black line is the Jordan-Wigner result in the thermodynamic limit with periodic boundary conditions. The other curves represent DMRG results for finite FM systems with open boundary conditions. Antiferromagnetic (AFM) $K$ yields the same energy. (b) Excitation gap in the thermodynamic limit from the Jordan-Wigner solution.}
\end{figure}

Unitary diagonalization of Eq.~\eqref{eq:ham:JW:BdG:matrix} 
yields a spectrum symmetric around zero, with energies $\{ \pm \epsilon_{k,n} \}$, $n=1,2$ given by
\begin{align}
\epsilon_{k,1}	&=	\sqrt{C_k-\sqrt{D_k}},	\quad \epsilon_{k,2}	=	\sqrt{C_k+\sqrt{D_k}},	\label{eq:JW:eigenvalues}
\end{align}
where
\begin{align}
C_k	&=	\left| A_k \right|^2	+ 	\left| P_k \right|^2	+	\left| Q_k \right|^2	=	4K^2	\left[	1	+ \cos\left(k\right) \sin^2\left(2\theta\right)\right]
\end{align}
and
\begin{align}
D_k	&=	\left( A_k^\star P_k + A_k P_k^\star \right)^2	-	\left( A_k^\star Q_k - A_k Q_k^\star \right)^2	\nonumber\\
&+	\left( P_k^\star Q_k + P_k Q_k^\star \right)^2	\\
&=	16K^4 \cos^2 \left(\frac{k}{2}\right)	\sin^2 \left( 2\theta\right)	\nonumber\\
&\left[ 3 + \cos \left( 4\theta\right) + 2\cos\left(k\right)\sin^2 \left(2\theta\right)\right].
\end{align}
$\epsilon_{k,1}$ and $\epsilon_{k,2}$ are called the acoustic and optical branches, respectively, in analogy with phonon terminology. Positive energy states represent physical excitations, while negative energy states stem from the redundancy in the description and are filled in the ground state, which has energy
\begin{align}
	E_0 &= -\frac{1}{2}\sum_k \left( \epsilon_{k,1} + \epsilon_{k,2}\right).	\label{eq:JW:GSenergy}
\end{align}
This function is plotted in black in Fig.~\ref{fig:energy:FM}(a).

Some important observations follow directly from the eigenvalues \eqref{eq:JW:eigenvalues}. First of all, the energies are independent of the sign of $K$. Second, since $\epsilon_{k,1}\leq \epsilon_{k,2}\quad \forall k,\theta$  the excitation gap is given by $\Delta \left( \theta\right)	=	2\min_k	\epsilon_{k,1} \left( \theta\right),$  which generically has extrema at $k=0,\pi$ and is plotted in Fig.~\ref{fig:energy:FM}(b). 
We note that the gap $\Delta(\theta)$ is best understood as the physical energy gap in the thermodynamic limit, i.e. the gap between a spontaneously $\mathbb{Z}_2$-symmetry-broken ground state and the first excited state above it. 
At finite system size, analysis of the gap in the Jordan-Wigner formalism requires careful treatment of boundary conditions and Bologiubov vacua \cite{PhysRevB.75.134415}, which is outside the scope of the current paper. In numerical calculations on finite-size systems the physical gap may be identified via $\Delta_2=E_2-E_0$, where $E_n$ is the $n$th lowest eigenvalue and multiplicity is taken into account.

In the Ising limits at $\theta=0,\,\pi/2$, the excitations are gapped, doubly degenerate and nondispersive, with $\epsilon_{k,1}=\epsilon_{k,2}=2\left| K\right|$ ($D_k=0$, $C_k=4K^2$). At the Kitaev point $\epsilon_{k,1} \left( \theta = \pi/4\right)	=	0	\quad	\forall k$, meaning that the excitations are nondispersive and gapless. The acoustic $+\epsilon_{k,1}$ branch thus becomes degenerate with the $-\epsilon_{k,1}$ branch, which  leads to a macroscopic degeneracy \cite{PhysRevB.89.104425}, and $h(k)$ becomes singular. As shown below, this degeneracy results in anomalous behavior at the Kitaev point. Away from the Ising and Kitaev limits, the excitations are dispersive and gapped.

\section{Numerical methods}\label{sec:methods}

We performed zero-temperature two-site DMRG \cite{PhysRevLett.69.2863, PhysRevB.48.10345} calculations using the DMRG++ software \cite{Alvarez2009} and open boundary conditions (OBC). The dynamical spin structure components $S^{aa}(k,\omega)$ were calculated in the Krylov correction-vector approach \cite{PhysRevB.60.335, PhysRevB.66.045114, PhysRevE.94.053308}, which works directly in frequency space and allows constant frequency resolution. The center-site approximation was employed, and elastic delta function peaks in $S^{zz}(k,\omega=0)$ due to static order away from the Kitaev point were removed by subtracting the ground state magnetization from the center-site operator; see the supplemental material for details \cite{Supplemental}. 
Since this procedure relies on a well-defined expectation value $\langle S_c^z\rangle \neq 0$ it is reliable only for a non-degenerate ground state. Thus, for $0 \leq \theta < \pi/4$ [$\pi/4 < \theta < \pi/2$] $S^{zz}(k,\omega)$ [$S^{xx}(k,\omega)$] was computed in the presence of a small uniform (staggered) symmetry-breaking magnetic pinning field of 
magnitude $10^{-6}\tilde{K}$ along $\hat{z}$ [$\hat{x}$] for FM (AFM) $K$, compatible with the static correlations; 
see Sec.~\ref{sec:results}.

Our main results (i.e. spectra) 
were obtained with $L=64$ sites, keeping up to $m_\mathrm{max}=1920$ states in the calculations. 
A Lorentzian broadening of $\eta=0.1\tilde{K}$ and a frequency step of $\Delta \omega = 0.025 \tilde{K}$ were also used. 
Truncation errors below $10^{-10}$ were targeted, which was easily achieved in practice (since most calculations used substantially fewer states than allowed by $m_\mathrm{max}$), except in the vicinity of the QCM, where the largest single truncation error was instead on the order of $10^{-6}$. We note that we obtained very similar results for the QCM also for a lower value $m_\mathrm{max}=1280$, albeit with a larger truncation error. 
Overall, calculations at the critical point dominated the computational effort; see the supplemental material \cite{Supplemental}, which also provides additional details for reproducing the numerical results.

We use a two-site unit cell as in Fig.~\ref{fig:zigzagchain} and designate momenta in units of the crystallographic lattice constant $1/c$. 
The momenta are labeled $k_n=2\pi n/N$, $n=0,1,\dots, N-1$, where $N=L/2$ is the number of unit cells. This effectively amounts to treating the system as if it were periodic, which introduces a minor error that vanishes in the thermodynamic limit. 
We use a Fourier transform convention that accounts for the position within the unit cell, which is taken to be $0$ for even sites and $c/2$ for odd sites. We note, however, that due a glide symmetry of Eq.~\eqref{eq:pauliham:morris} (composed of translation by $\tilde{c}=c/2$ and a spin flip) the resulting spin structure factors are insensitive to the unit cell doubling and show periodicity by $2\pi/\tilde{c} = 4\pi/c$ \cite{doi:10.1073/pnas.2007986117}. As such, the results can readily be reinterpreted for a single-site unit cell by scaling $k$.

\section{Results}\label{sec:results}
\begin{figure}
	\includegraphics[width=\columnwidth]{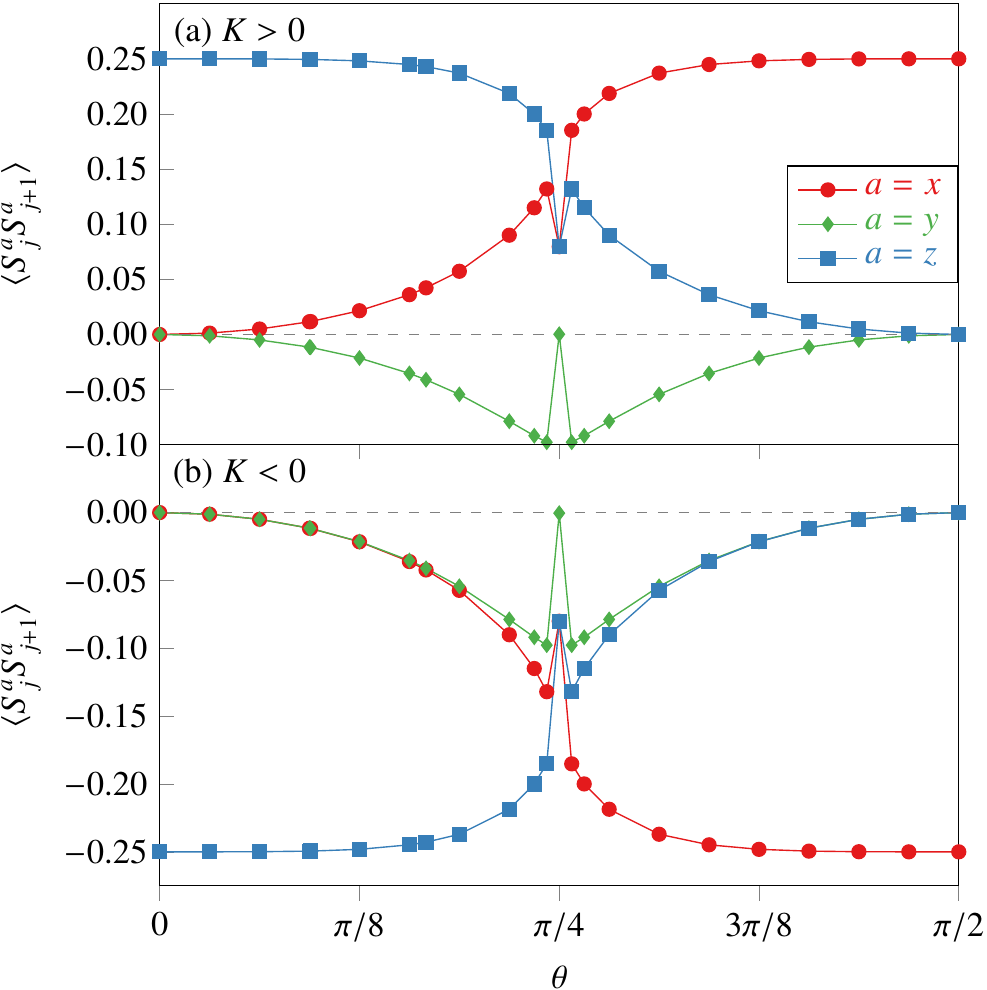}
	\caption{\label{fig:statics}Nearest-neighbor static spin-spin correlation functions for (a) $K>0$ and (b) $K<0$. All data is for $L=64$ and $m_\mathrm{max}=1280$.}
\end{figure}
Figure~\ref{fig:energy:FM}(a) shows the groundstate energy from DMRG and from the continuum limit of Eq.~\eqref{eq:JW:GSenergy}. The numerical results indicate quick convergence towards the exact result with system size $L$. For $\theta$ away from $\theta_c=\pi/4$ very large system sizes can be reached. 
Figure~\ref{fig:statics} shows static groundstate nearest-neighbor correlation functions from the DMRG calculations (with zero pinning fields). Four sites at each end of the chain were neglected to minimize boundary effects, such that the nearest-neighbor correlations were averaged over the interior $L-2\times 4 -1$ bonds.  In both the FM [Fig.~\ref{fig:statics}(a)] and AFM [Fig.~\ref{fig:statics}(b)] cases the system is characterized by large $\left| \langle S_j^zS_{j+1}^z\rangle\right|$ for $0\leq \theta < \frac{\pi}{4}$ and large $\left| \langle S_j^xS_{j+1}^x\rangle\right|$ for $\pi/4< \theta \leq \pi/2$, reflecting the change of the easy axis. At the Kitaev point all correlation functions decrease, associated with a disordered state, as previously discussed in Ref.~\cite{PhysRevB.89.104425}.

\begin{figure}
	\includegraphics[width=\columnwidth]{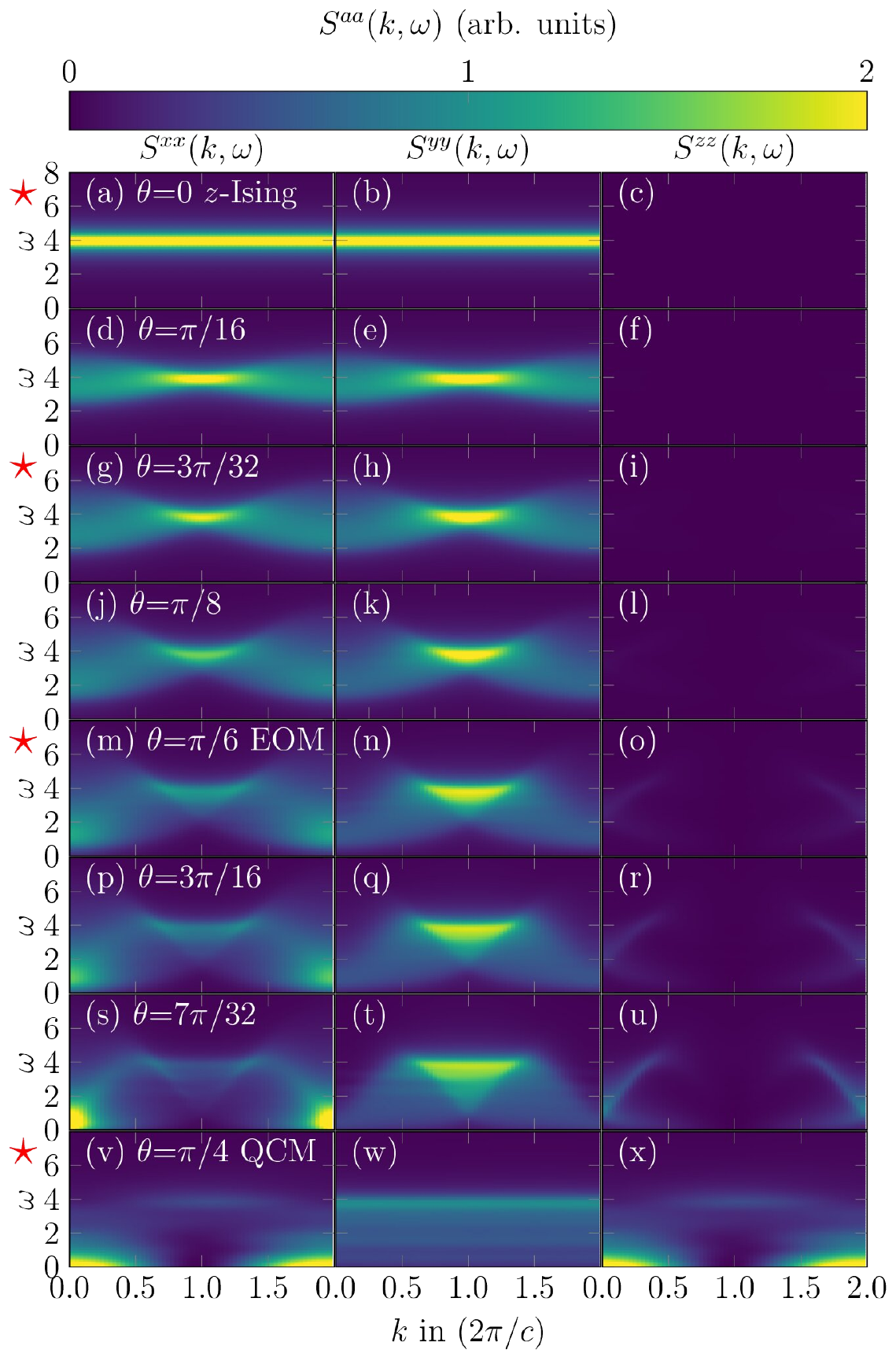}
	\caption{\label{fig:spectra:FM}Dynamical spin structure factor for the 1D FM GCM for different values of $\theta$. Red stars indicate notable special cases: (a)-(c) show the FM Ising chain, where spin waves are non-dispersive and the inelastic weight is concentrated in transverse scattering. (g)-(i) show results for $\theta\approx 17^\circ$, which Ref.~\cite{Morris2021} proposed is relevant to CoNb$_2$O$_6$. (m)-(o) show spectra for the FM $e_g$-orbital model (EOM), and (v)-(x) spectra for the Kitaev spin chain or QCM. Energies are given in units of $|K|$. Other panels show spectra for intermediate values of $\theta$. As $\theta$ increases from $0^\circ$, the excitations become dispersive and the spin gap gradually decreases until it closes at $\theta=45^\circ$, where the nature of the scattering changes. All results shown were obtained for $L=64$ sites and OBC. Elastic $\delta$-function peaks in $S^{zz}(k,\omega)$ were removed for $\theta<\pi/4$; see Sec.~\ref{sec:methods}.
	}
\end{figure}

Figure~\ref{fig:spectra:FM} shows the diagonal components of the dynamical spin structure factor for ferromagnetic $K>0$ and various values of the angle $0\leq \theta \leq \frac{\pi}{4}$ in the Ising-like coordinate system. 
We use units in which $|K|=1$. The range includes the Ising and Kitaev limits at the ends of the interval, as well as $\theta=3\pi/32\approx 17^\circ$ and $\theta=\pi/6=30^\circ$, corresponding to 
proposed values for CoNb$_2$O$_6$ \cite{Morris2021} and the FM $e_g$-orbital model \cite{PhysRevB.89.104425}, respectively. Spectra for the range $\frac{\pi}{4}\leq \theta \leq \frac{\pi}{2}$ are related to those shown by the substitution $\theta\rightarrow \pi/2 - \theta,$ $x\leftrightarrow z$.

At $\theta=0$ in Fig.~\ref{fig:spectra:FM}(a)-(c), we have a FM $z$-Ising chain with gapped, non-dispersive excitations. In this limit the ground state has the form $c_\uparrow \left| \uparrow\uparrow\uparrow \dots \right\rangle+c_\downarrow \left| \downarrow\downarrow\downarrow \dots \right\rangle$, so $S^{zz}(k,\omega)\propto \delta(k)\delta(\omega)$ becomes trivial. As discussed in Sec.~\ref{sec:methods} this elastic peak was subtracted from the plotted spectrum. The true inelastic scattering is contained purely in the transverse components. These probe the energy related to domain walls, which have energy $4|K|$.

For $0<\theta<\pi/4$ the presence of additional terms in the Hamiltonian induces domain wall motion \cite{Morris2021}, which translates into dispersive excitations and scattering continua in the transverse components. This simple physical picture is familiar from FM XY and XXZ chains, but also holds here in the presence of a site-alternating $\Gamma$ term. Initially, as in Fig.~\ref{fig:spectra:FM}(d)-(e), the $S^{xx}(k,\omega)$ and $S^{yy}(k,\omega)$ components appear fairly symmetric, both in their bow-tie-like shape and spectral distribution, which has most weight near $\omega=4|K|$ and $k=2\pi/c$. However, as $\theta$ is increased, the spectral weight in $S^{xx}(k,\omega)$ is redistributed towards the $\Gamma$ point; see Fig.~\ref{fig:spectra:FM}(m),(p),(s). At the same time, the delta function peak in $S^{zz}(k,\omega)$ becomes less dominant and some dispersive inelastic scattering becomes visible in Fig.~\ref{fig:spectra:FM}(o),(r),(u). As $\theta\rightarrow \pi/4$ the spin excitations become gapless as predicted by the Jordan-Wigner solution, with significant weight at $\omega=0$ in $S^{xx}(k,\omega)$ and $S^{zz}(k,\omega)$, while $S^{yy}(k,\omega)$ becomes more diffuse and completely flat with a concentration of spectral weight along the top of the spectrum; see Fig.~\ref{fig:spectra:FM}(w). 

\begin{figure}
	\includegraphics[width=\columnwidth]{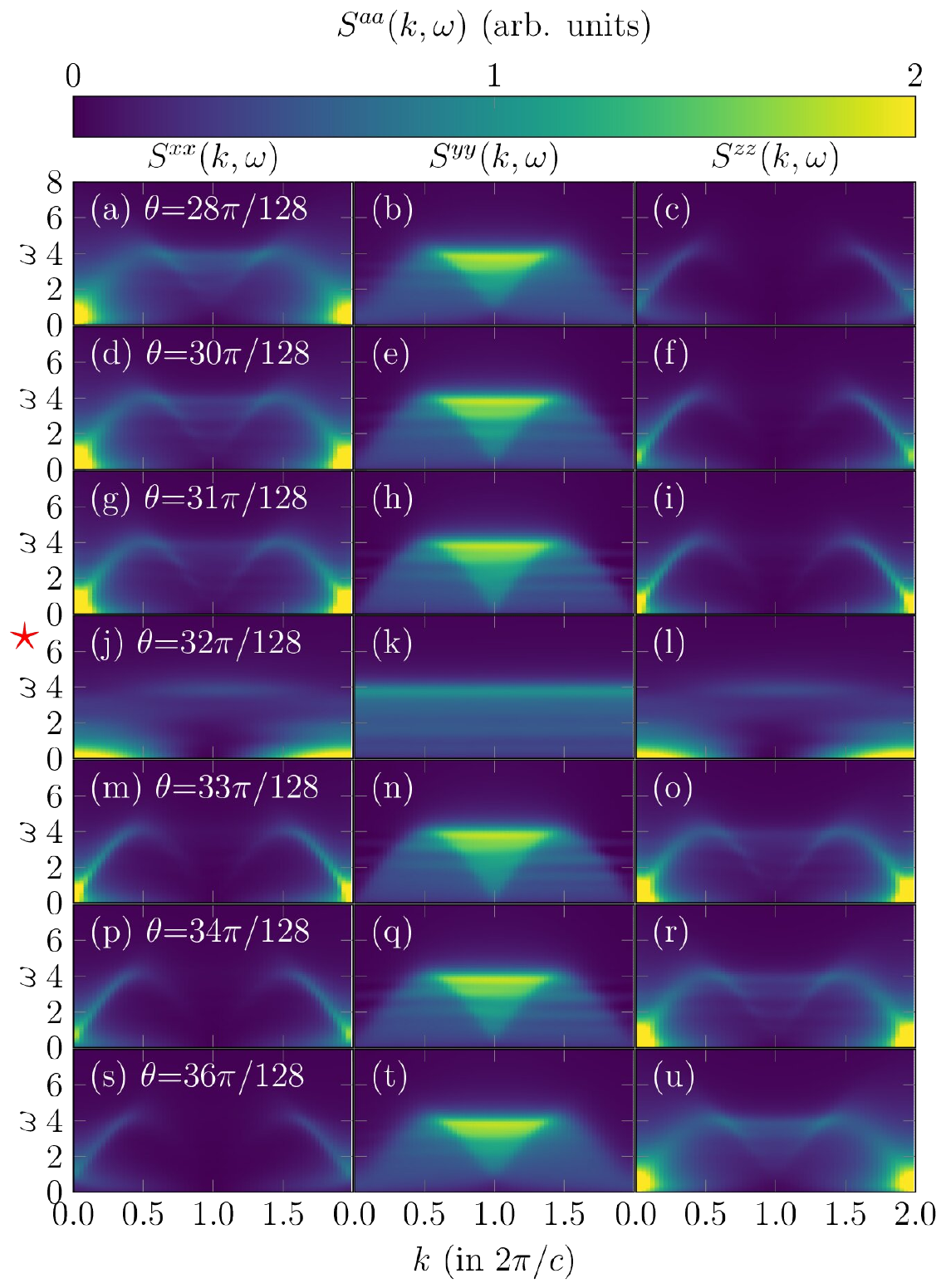}
	\caption{\label{fig:spectra:FM:nearKitaev}Dynamical spin structure factor for the 1D FM GCM for different values of $\theta$ close to the Kitaev limit $\theta=\pi/4$ [panels (j)-(l), marked with the red star]. Although the $S^{xx}(k,\omega)$ and $S^{zz}(k,\omega)$ change reasonably smoothly across the critical point, there is a sudden change in the $S^{yy}(k,\omega)$ component. All results shown were obtained for $L=64$ sites and OBC. Elastic $\delta$-function peaks in $S^{xx/zz}(k,\omega)$ were removed for $\theta\neq \pi/4$; see Sec.~\ref{sec:methods}.}
\end{figure}
This highly unusual dispersionless scattering feature appears very suddenly at the critical point. To see just how abrupt the spectra change we consider additional values of $\theta$ close to $\theta_c$ in Fig.~\ref{fig:spectra:FM:nearKitaev}. The qualitative form of $S^{yy}(k,\omega)$ is symmetric around $\theta=\pi/4$, and unchanged in the $28\pi/128 \leq \theta \leq 31\pi/128$ range, yet suddenly changes at the gap closing point. 
Given the abruptness, one may be tempted to ask if the spectrum in Fig.~\ref{fig:spectra:FM}(w) / Fig.~\ref{fig:spectra:FM:nearKitaev}(k) is correct. We stress that, although the Kitaev point is the most computationally challenging, this spectrum is \emph{not} a simple numerical artifact. Instead, the anomalous behavior is linked directly to the extensive ground state degeneracy and restructuring of the Hilbert space seen in the Jordan-Wigner solution. 
From the analytical results of Perk \emph{et al.} \cite{Perk1977} for time-dependent correlations we have obtained the structure of $S^{yy}(k,\omega)$ at $\theta=\pi/4$. It features a $k$-independent continuum for $0\leq \omega \leq 4|K|$ with divergent intensity towards the top of the spectrum, in agreement with the numerical result. See  Appendix~\ref{app:dispersionlesscontinuum} for details of the derivation. 
We also note that, although the system at $\theta=\pi/4$ is referred to as a Kitaev spin chain, the behavior in the isotropic Kitaev honeycomb model is markedly different. That model realizes a quantum spin liquid with gapless Majorana excitations, yet remarkably its spin excitation spectrum remains gapped \cite{PhysRevLett.112.207203}. The gap is related to an emergent static gauge field \cite{PhysRevLett.112.207203}, which is absent in the chain \cite{PhysRevLett.98.087204}.

\begin{figure}
	\includegraphics[width=\columnwidth]{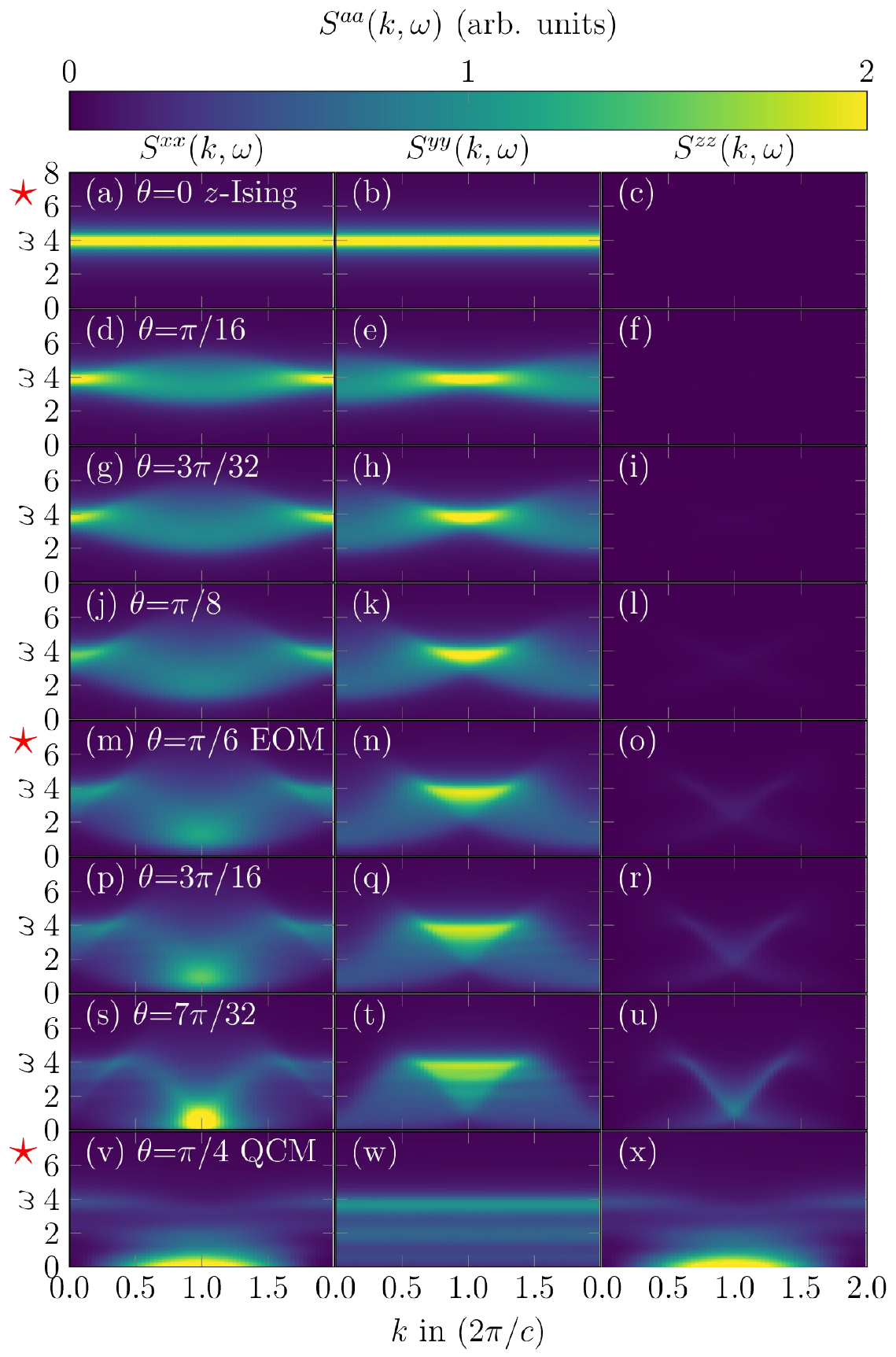}
	\caption{\label{fig:spectra:AFM}Spectra for the 1D AFM GCM for different values of $\theta$. Red stars indicate the AFM Ising chain (a)-(c), the AFM $e_g$-orbital model (m)-(o), and the AFM QCM (v)-(x).  Other panels show spectra for intermediate values of $\theta$. All results shown were obtained for $L=64$ sites and OBC.  Elastic $\delta$-function peaks in $S^{zz}(k,\omega)$ were removed for $\theta<\pi/4$; see Sec.~\ref{sec:methods}.
	}
\end{figure}
The antiferromagnetic case in Fig.~\ref{fig:spectra:AFM} shows the same behavior in the transverse $S^{yy}(k,\omega)$ component, however the $S^{xx}(k,\omega)$ and $S^{zz}(k,\omega)$ components are modified compared to the FM case. This is due to a canonical transformation where spins on one sublattice (e.g. even sites) are rotated by $\pi$ around $\hat{y}$, taking $H_1\rightarrow -H_1$. For the dynamics it implies a $2\pi/c$ shift in $k$ for $S^{xx/zz}(k,\omega)$ between the FM and AFM cases. The most apparent consequence 
is the shift of spectral weight in $S^{zz}(k,\omega)$ from $k=0$ to $k=2\pi/c$, reflecting N\'eel correlations. Its origin is also clear 
from the AFM state in the Ising limit, $c_\uparrow \left| \uparrow\downarrow\uparrow \dots \right\rangle+c_\downarrow \left| \downarrow\uparrow\downarrow \dots \right\rangle$, yielding $S^{zz}(k,\omega)\propto \delta(k-2\pi/c)\delta(\omega)$. As other terms are introduced in the Hamiltonian, a continuum develops in the transverse components, reminiscent of the AFM XXZ chain \cite{10.1143/PTP.63.743}. 
A qualitative difference compared to the FM case is that the bow-tie-like shapes of $S^{xx}(k,\omega)$ for low $\theta$ are replaced by more rounded shapes [compare, for example, Fig.~\ref{fig:spectra:FM}(g) and Fig.~\ref{fig:spectra:AFM}(g)], which follows from the $2\pi/c$ shift. Essentially, both shapes can be understood as emerging from the dispersionless excitations in the Ising limit by gradually shifting spectral intensity towards $k=0$ or $k=2\pi/c$ with increasing $\theta$. Interestingly, in both the FM and AFM cases, the Ising limit scattering leaves strong imprints on the spectra at finite $\theta$, whose $k=2\pi/c$ and $k=0$ excitations, respectively, retain their energy scale.

\section{Discussion and conclusion}\label{sec:discussion}
The lack of U(1) symmetry around the easy-axis in Eq.~\eqref{eq:pauliham:morris} implies generally that the two transverse components of the dynamical spin structure factor will differ. This is seen in Figs.~\ref{fig:spectra:FM}, \ref{fig:spectra:AFM} for $\theta$ large enough, where $S^{xx}(k,\omega)\neq S^{yy}(k,\omega)$. 
In the ferromagnetic case and for low $\theta$, however, 
$S^{xx}(k,\omega)\approx S^{yy}(k,\omega)$ is a good approximation. We note that this assumption was made in the analysis of inelastic neutron scattering data on CoNb$_2$O$_6$ in Ref.~\cite{Coldea2010}. For $\theta=3\pi/32$ [see Figs.~\ref{fig:spectra:FM}(g),(h)] we find that $S^{yy}(k,\omega)$ has a $\approx 15\%$ higher peak intensity than $S^{xx}(k,\omega)$, but essentially the same integrated spectral weight. Given that Figs.~\ref{fig:spectra:FM}(g),(h) also indicate an approximately symmetric distribution of the spectral weight, we conclude that the assumption is justified also under the Hamiltonian parameters proposed in Ref.~\cite{Morris2021}. However, for systems approximately described by Eq.~\eqref{eq:pauliham:morris} at higher $\theta$ or AFM $K<0$, spin-polarization-resolved spectroscopic experiments would be preferable and provide important information about the bond directionality of interactions.

In systems of weakly coupled Ising chains the interchain effects can be incorporated through an effective longitudinal magnetic field that becomes non-zero in the ordered phase \cite{Coldea2010, PhysRevLett.90.177206}. According to the proposal of Ref.~\cite{Morris2021}, long-distance properties of CoNb$_2$O$_6$, such as the THz spectrum, can be well described by $H = H_1 - h_z \sum_i S_i^z$ with $K=0.57$meV and $h_z=0.04$ meV. However, we have found this model insufficient to reproduce short-distance features seen in the INS data of Ref.~\cite{Coldea2010}, in particular it fails to reproduce the upwards curvature of the dispersion at $k=2\pi/c$. More realistic spin models for this material feature additional interactions, notably including a second-nearest neighbor AFM Ising interaction \cite{Coldea2010, PhysRevB.83.020407, doi:10.1073/pnas.2007986117} which appears to be necessary for a full description of the material in the entire Brillouin zone. 

Beyond CoNb$_2$O$_6$, we note that bond-dependent interactions are inherently related to the geometry of electron orbitals and hopping paths. This means that, except in fine-tuned systems, one generally expects that additional symmetry-allowed spin interaction terms may be present, much like is seen in the honeycomb Kitaev candidates \cite{PhysRevLett.112.077204}. In materials with well-separated chains, the impact of such terms can likely be tuned or minimized using pressure or strain. Some such terms could potentially also help stabilize the region of the disordered phase of the QCM, which otherwise occupies a singular point in the phase diagram. The interchain coupling itself can have important effects on, e.g., magnetic order. However, as long as it is weak it often does not significantly modify the high-energy spin dynamics, which can remain effectively one-dimensional above some cut-off frequency. Thus, there is hope of realizing a proximate 1D QCM, and more generally chain systems with substantial bond-dependent interactions.

Here, we have studied the dynamical spin structure factor of the spin-$1/2$ generalized compass chain, as a function of the angle between the local Ising directions. We find smooth changes in the components in the plane spanned by the Ising directions, but a sudden change in the perpendicular component at the Kitaev point. This is one of several anomalies that stem from the closing of the excitation gap and the development of an extensive groundstate degeneracy. Our results can help guide the interpretation and design of spectroscopic experiments on materials with similar bond-dependent interactions. Future work may extend the analysis to chains with additional symmetry-allowed interactions, ladder models \cite{PhysRevB.80.014405}, or chains in the presence of magnetic fields in which additional quantum phase transitions and also interesting soliton physics have been reported \cite{PhysRevResearch.5.L012027}.

\begin{acknowledgments}
	We thank J.~H.~H. Perk for helpful comments and C.~D. Batista for useful discussions. 
	The work of PL and ED was supported by the U.S. Department of Energy, Office of Science, Basic Energy Sciences, Materials Sciences and Engineering Division. G.A. was supported in part by the Scientific Discovery through Advanced Computing (SciDAC) program funded by the U.S. DOE, Office of Science, Advanced Scientific Computing Research and BES, Division of Materials Sciences and Engineering.
\end{acknowledgments}

\appendix
\section{Dispersionless continuum at the Kitaev point}\label{app:dispersionlesscontinuum}
Perk \emph{et al}. \cite{Perk1977} studied time-dependent correlations in the inhomogeneous one-dimensional XY-model with alternating interactions,
\begin{align}
	H	&=	2\sum_{i=1}^L \left[ J_i^x \tilde{S}_i^x \tilde{S}_{i+1}^x + J_i^y \tilde{S}_i^y \tilde{S}_{i+1}^y \right],	\label{eq:XYmodel}
\end{align}
where
\begin{align}
	J_{2i}^x = J_\mathrm{e}^x,\, J_{2i}^y = J_\mathrm{e}^y	&\quad\text{and}\quad	J_{2i+1}^x = J_\mathrm{o}^x,\, J_{2i+1}^y = J_\mathrm{o}^y.
\end{align}
The subscripts $\mathrm{e}$ and $\mathrm{o}$ denote even and odd, respectively. Here we have written the Hamiltonian in terms of $\tilde{S}$ operators to indicate the same coordinate system as was used in $H_3$, Eq.~\eqref{eq:spinham:rotated}. In the Kitaev limit, $\theta=\pi/4$ in Eq.~\eqref{eq:spinham:rotated}, and
\begin{equation}
	J_\mathrm{e}^x = J_\mathrm{o}^y	\equiv J,	\quad J_\mathrm{o}^x = J_\mathrm{e}^y = 0,
\end{equation}
in Eq.~\eqref{eq:XYmodel}, we identify $\tilde{K}=-2J$. At this isolated parameter point we can make use of the analytical results for the real-space and real-time dependent correlation function $\langle \tilde{S}_i^z (t) \tilde{S}^{z}_{i+r}(0)\rangle$ or the intermediate scattering function $\tilde{I}^{zz}(k,t)$ in Eqs. (4.20) and (4.26) of Ref.~\cite{Perk1977}. Due to the transformation \eqref{eq:spinrot}, these correlations are equivalent to $yy$ correlations in the Ising-like coordinate system of $H_1$ in Eq.~\eqref{eq:pauliham:morris}.

Taking the Kitaev and zero-temperature limits, one finds their Eq.~(4.26) simplifies substantially to
\begin{align}
	I^{yy}(k,t)	&=	\frac{1}{8\pi} \int_0^{2\pi} \mathrm{d}\varphi \exp \left[-i\Lambda_1\left(\varphi\right) t \right],
\end{align}
where
\begin{align}
	\Lambda_1\left(\varphi \right)	&=	\sqrt{2} \left| J \right| \sqrt{1-\cos \left( 2\varphi\right)}.
\end{align}
Note that there is no $k$-dependence in this limit. Next, the Fourier transform to frequency space yields
\begin{align}
	S^{yy}\left( k,\omega\right)	&\propto \frac{1}{2\pi}\int_{-\infty}^\infty \mathrm{d}t e^{i\omega t} I^{yy} \left( k, t\right)	\nonumber\\
									&= \frac{1}{16\pi^2}	\int_0^{2\pi} \mathrm{d}\varphi \int_{-\infty}^\infty \mathrm{d}t \exp \left[ it \left( \omega - 2\left| J \right| \left| \sin \varphi \right| \right)\right]	\nonumber\\
									&=	\frac{1}{8\pi} \int_0^{2\pi} \mathrm{d}\varphi \, \delta \left( \omega - 2 \left| J \right| \left| \sin \varphi \right| \right)	\nonumber\\
									&=	\frac{1}{4\pi} \int_0^{\pi} \mathrm{d}\varphi \, \delta \left( \omega - 2 \left| J \right| \sin \varphi \right),	\label{eq:analytical:skomega}
\end{align}
where the $\delta$-function produces a continuum. The last step in Eq.~\eqref{eq:analytical:skomega} makes its argument continuously differentiable in $\varphi$ such that the composition property of the $\delta$-function can be used. There are three different cases: (i) $\omega<0$ or $\omega>2\left| J\right|$, (ii) $0\leq \omega < 2\left| J \right|$, and (iii) $\omega=2\left| J \right|$. In the trivial case (i) the function $g(\varphi)=\omega - 2\left| J \right| \sin\varphi$ has no roots, making $S(k,\omega)$ vanish. In case (ii) there are two roots in the interval $[0,\pi]$,
\begin{align}
	\varphi_1 &= \sin^{-1} \left(\frac{\omega}{2\left| J \right|}\right),\quad \varphi_2 = \pi - \sin^{-1} \left(\frac{\omega}{2\left| J \right|}\right).
\end{align}
Both roots leave $g^\prime \left( \varphi_n\right)\neq 0$, making $S(k,\omega)$ finite throughout the entire frequency range. Finally, in case (iii) there is only one unique root, $\varphi_3=\pi/2$. Since $g^\prime\left(\pi/2\right)=0$ it follows that $S\left(k,\omega=2\left| J \right|\right)$ diverges. This is consistent with what we observe numerically in Figs. \ref{fig:spectra:FM}(w), \ref{fig:spectra:AFM}(w), where the intensity is found to be concentrated along the top edge of the spectrum, with a weaker dispersionless continuum below it. The lack of a sharp divergence at $\omega=2\left| J \right| = 4\left| K\right|$ in the numerical result is due to the Lorentzian frequency broadening.

%

%
	
\end{document}


\title{Supplemental material for \texorpdfstring{\\}{}``Spin dynamics in the generalized quantum spin compass chain''}
\author{Pontus Laurell}
\affiliation{Department of Physics and Astronomy, University of Tennessee, Knoxville, Tennessee 37996, USA}
\affiliation{Computational Sciences and Engineering Division, Oak Ridge National Laboratory, Oak Ridge, Tennessee 37831, USA}
\author{Gonzalo Alvarez}
\affiliation{Computational Sciences and Engineering Division, Oak Ridge National Laboratory, Oak Ridge, Tennessee 37831, USA}
\author{Elbio Dagotto}
\affiliation{Department of Physics and Astronomy, University of Tennessee, Knoxville, Tennessee 37996, USA}
\affiliation{Materials Science and Technology Division, Oak Ridge National Laboratory, Oak Ridge, Tennessee 37831, USA}

\date{\today}
\maketitle

\section{Supplemental: Reproducing the DMRG results.}
Here we provide detailed instructions on how to reproduce the DMRG results used in the main text. The results reported in this work were obtained with the DMRG++ version 6.05 and PsimagLite version 3.04.

The \textsc{DMRG}++ computer program 
can be obtained with:
\begin{verbatim}
git clone https://github.com/g1257/dmrgpp.git
\end{verbatim}
Dependencies include the BOOST and HDF5 libraries, and PsimagLite. The latter can be obtained with:
\begin{verbatim}
git clone https://github.com/g1257/PsimagLite.git
\end{verbatim}
To compile:
\begin{verbatim}
cd PsimagLite/lib; perl configure.pl; make
cd ../../dmrgpp/src; perl configure.pl; make
\end{verbatim}
To simplify commands below we also run
\begin{verbatim}
export PATH="<PATH-TO-DMRG++>/src:$PATH"
export SCRIPTS="<PATH-TO-DMRG++>/scripts"
\end{verbatim}

The documentation can be found at 
\nolinkurl{https://g1257.github.io/dmrgPlusPlus/manual.html} or can be obtained
by doing 
\begin{verbatim}cd ../..; git clone https://github.com/g1257/thesis.git; cd dmrgpp/doc; ln -s ../../thesis/thesis.bib; 
make manual.pdf
\end{verbatim}

\subsection{Ground states}

The ground state is obtained by running DMRG++ with an appropriate input file, e.g. \texttt{dmrg -f inputGS.ain}, where a typical input (for $L=64$ sites, $\theta=\pi/4$, ferromagnetic case) in the Ising-like coordinate system is sketched below:
\begin{verbatim}
##Ainur1.0
TotalNumberOfSites=64;
NumberOfTerms=4;

MagneticFieldX=[0.0,...];
MagneticFieldZ=[0.0,...];

### SxSx
gt0:DegreesOfFreedom=1;
gt0:GeometryKind="chain";
gt0:GeometryOptions="ConstantValues";
gt0:dir0:Connectors=[-0.5];

### SySy
gt1:DegreesOfFreedom=1;
gt1:GeometryKind="chain";
gt1:GeometryOptions="ConstantValues";
gt1:dir0:Connectors=[0.00];

### SzSz
gt2:DegreesOfFreedom=1;
gt2:GeometryKind="chain";
gt2:GeometryOptions="ConstantValues";
gt2:dir0:Connectors=[-0.5];

### SxSz+SzSx
gt3:DegreesOfFreedom=1;
gt3:GeometryKind="chain";
gt3:GeometryOptions="none";
gt3:dir0:Connectors=[0.5, -0.5, ... ];

Model="KitaevWithGammasReal";
HeisenbergTwiceS=1;

SolverOptions="twositedmrg,calcAndPrintEntropies";
Version="twistedkitaev";
InfiniteLoopKeptStates=1280;
FiniteLoops=[
[ 31, 1280, 8],
[-62, 1280, 8],
[ 62, 1280, 2],
[-62, 1280, 3]];

# Keep a maximum of m states, but allow truncation with tolerance and minimum states as
TruncationTolerance="1e-10,100";

# Tolerance for Lanczos
LanczosEps=1e-10;
int LanczosSteps=600;
\end{verbatim}
where $\dots$ in the \texttt{gt3:dir0:Connectors} needs to be replaced by the repeating pattern for all $L-1$ bonds of the open system. The \texttt{KitaevWithGammasReal} model used above was implemented for this project as a special case of the \texttt{KitaevWithGammas} model (which implements also $S^xS^y+S^yS^x$ and $S^yS^z+S^zS^y$ interactions) in order to treat the Hamiltonian as real-valued, substantially reducing memory and time requirements.

\begin{figure}
	\includegraphics[width=0.6\columnwidth]{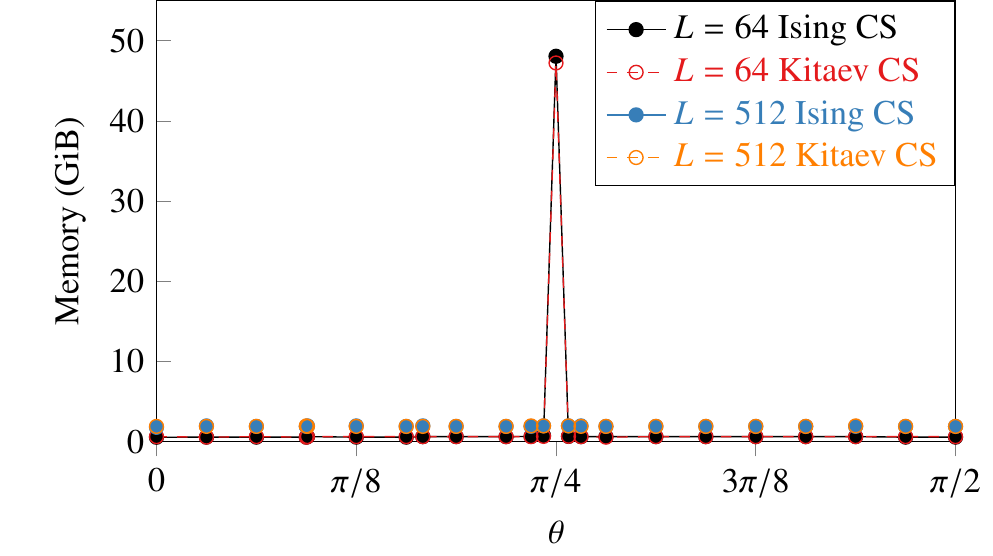}
	\caption{\label{fig:supp:memory}
		Memory usage in ground state runs as function of $\theta$ for $K>0$. For $L=64$ up to $m=1280$ states were kept in the calculation. Higher $m$ was allowed for $L=512$, but not required for the shown values of $\theta \neq \pi/4$. $L=128$ runs using up to $m=2560$ (not shown) were also performed, and were found to use up to $\approx 290$ GiB at $\theta=\pi/4$.
	}
\end{figure}

\begin{figure}
	\includegraphics[width=0.6\columnwidth]{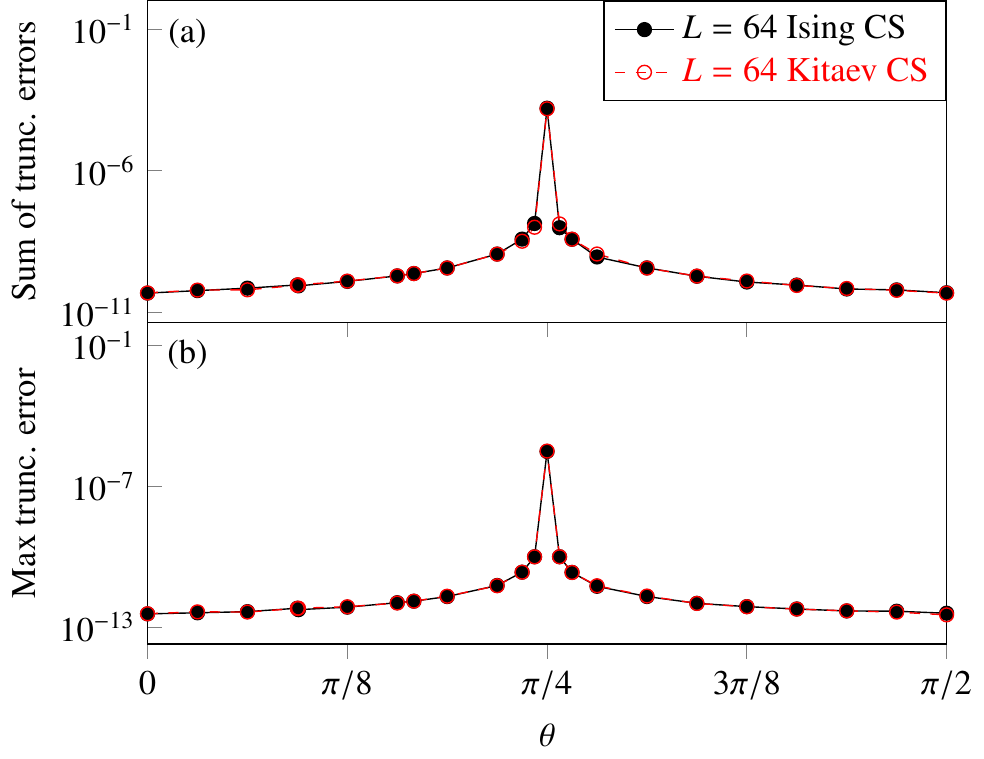}
	\caption{\label{fig:supp:truncation}
		Truncation errors as function of $\theta$ for $K>0$, $L=64$ and $m_\mathrm{max}=1280$. (a) Sum of truncation errors across all sweeps and (b) maximal single truncation error in DMRG ground state runs. The macroscopic degeneracy and criticality found at $\theta=\pi/4$ requires a high number of kept DMRG states (or bond dimension).
	}
\end{figure}

As mentioned in the main text, the Kitaev point $\theta=\pi/4$ is particularly computationally demanding. The difference compared to other values of $\theta$ may be seen clearly in the memory usage shown in Fig.~\ref{fig:supp:memory} and the behavior of the truncation errors in Fig.~\ref{fig:supp:truncation}. These figures show only the FM case ($K>0$), but the behavior in the AFM case is very similar. In both figures Ising and Kitaev CS refers to the Ising-like and Kitaev-like coordinate systems introduced in the main text, which perform nearly identically.

\newpage
\subsection{Obtaining zero-temperature spectra}

The next step is to calculate dynamics, using the saved ground state as an input. It is convenient to do the dynamics run in a subdirectory \texttt{Sab}, where $a,b\in \{x,y,z\}$. For example, for the $S^{zz}(k,\omega)$ component do \verb!cp inputGS.ain Szz/inputSzz.ado! and add/modify the following lines in \texttt{inputSzz.ado},
\begin{verbatim}
SolverOptions="twositedmrg,restart,minimizeDisk,CorrectionVectorTargeting";

### The finite loops pick up where gs run ended! I.e. the edge.
FiniteLoops=[
[ 62, 1280, 2],
[-62, 1280, 2]];

# RestartFilename is the name of the GS .hd5 file (extension is not needed)
string RestartFilename="../inputGS";

# The weight of the g.s. in the density matrix
GsWeight=0.1;

# Legacy thing, set to 0
real CorrectionA=0;

# Fermion spectra has sign changes in denominator. For boson operators (as in here) set it to 0
integer DynamicDmrgType=0;

# The site(s) where to apply the operator below. Here it is the center site.
TSPSites=[32];

# The delay in loop units before applying the operator. Set to 0 for all restarts to avoid delays.
TSPLoops=[0];

# If more than one operator is to be applied, how they should be combined.
# Irrelevant if only one operator is applied, as is the case here.
TSPProductOrSum="sum";

# How the operator to be applied will be specified
string TSPOp0:TSPOperator="expression";

# The operator expression
string TSPOp0:OperatorExpression="sz";

# Apply operator to ground state
string TSPApplyTo="|X0>";

# How is the freq. given in the denominator (Matsubara is the other option)
CorrectionVectorFreqType="Real";

# This is a dollarized input, so the omega will change from input to input.
CorrectionVectorOmega=$omega;

# The broadening for the spectrum in omega + i*eta
real CorrectionVectorEta=0.10;

# The algorithm
CorrectionVectorAlgorithm="Krylov";

#The labels below are ONLY read by manyOmegas.pl script

# How many inputs files to create
#OmegaTotal=101

# Which one is the first omega value
#OmegaBegin=0.0

# Which is the "step" in omega
#OmegaStep=0.025

# Because the script will also be creating the batches, indicate what to measure in the batches
#Observable=sz
\end{verbatim}
For other components $S^{ab}(k,\omega)$, modify the lines \texttt{\#Observable=sz} and \texttt{OperatorExpression="sz"} according to the appropriate $S^a$ and $S^b$ operators. Note that, in the real-valued run mode of the \texttt{KitaevWithGammasReal} model the only defined operators are the real-valued \texttt{sx}${}:=S^x$, \texttt{sybar}${}:=iS^y$, \texttt{sz}${}:=S^z$ (as well as \texttt{splus}${}:=S^+$, \texttt{sminus}${}:=S^-$).  These operators are sufficient for all diagonal components $S^{aa}(k,\omega)$ and the off-diagonal $S^{xz/zx}(k,\omega)$ components. [For $S^{yy}(k,\omega)$ one may use $S^a={}$\texttt{sybar}, $S^b={}$\texttt{(-1.0)*sybar}.] Off-diagonal elements involving a single $S^y$ are supported by adding the \texttt{useComplex} to the \texttt{SolverOptions} list, which enables directly using the \texttt{sy}${}=S^y$ operator.

Naively proceeding as above results in delta function peaks at zero momentum and frequency in $S^{xx}(k,\omega)$ and $S^{zz}(k,\omega)$ when far enough away from $\theta=\pi/4$. These can, in principle, be resolved along with the inelastic scattering in the correction vector method with sufficient number of states. However, in practice we found that the peaks completely overshadowed the inelastic scattering at several values of $\theta$. To avoid this issue, we subtract the $\langle S_c^x\rangle$ and $\langle S_c^z\rangle$ ground state magnetizations from the center-site operator in the affected components of $S^{aa}(k,\omega)$, e.g. by writing 
\begin{verbatim}
	string TSPOp0:OperatorExpression="sz+(.49627907680)*identity";
\end{verbatim}
for $S^{zz}(k,\omega)$, $\theta=\pi/8$ and $L=128$. 
This procedure relies on $\langle S^a_c\rangle$ being well-defined, which is guaranteed when the groundstate is nondegenerate. In practice, in DMRG calculations on systems with open boundaries this is often the case simply due to the finite-size gap. In the presence of symmetries, however, DMRG can select an approximate groundstate that is an arbitrary linear combination of the symmetry-related ground states. We found that this occurred in our study, and often found it necessary to introduce a small symmetry breaking magnetic field in both GS and dynamics runs, e.g.
\begin{verbatim}
	MagneticFieldZ=[0.000001,...];
\end{verbatim}
in the FM case, or 
\begin{verbatim}
	MagneticFieldZ=[-0.000001, 0.000001,...];
\end{verbatim}
in the AFM case. In the latter case, the pattern must be repeated manually for all $L$ sites.

Then, one input per $\omega$ can be generated and submitted using the \texttt{manyOmegas.pl} script:
\begin{verbatim}
perl -I ${SCRIPTS} ${SCRIPTS}/manyOmegas.pl inputSzz.ado BatchTemplate <test/submit>.
\end{verbatim}
It is recommended to run with \texttt{test} first to verify correctness, before running with \texttt{submit}. Depending on the machine and scheduler, the BatchTemplate can be e.g. a PBS script.  The key is that it contains a line
\begin{verbatim}
dmrg -f $$input "<X0|$$obs|P1>,<X0|$$obs|P2>,<X0|$$obs|P3>" -p 10
\end{verbatim}
which allows \texttt{manyOmegas.pl} to fill in the appropriate input for each generated job batch. After all outputs have been generated,
\begin{verbatim}
perl -I ${SCRIPTS} ${SCRIPTS}/procOmegas.pl -f inputSzz.ado -p
perl ${SCRIPTS}/pgfplot.pl
\end{verbatim}
can be used to process and plot the results.